\documentstyle[12pt,epsf]{article}
\begin{document}
\catcode`@=11
\long\def\@caption#1[#2]#3{\par\addcontentsline{\csname
  ext@#1\endcsname}{#1}{\protect\numberline{\csname
  the#1\endcsname}{\ignorespaces #2}}\begingroup
    \small
    \@parboxrestore
    \@makecaption{\csname fnum@#1\endcsname}{\ignorespaces
#3}\par
  \endgroup}
\catcode`@=12
\def\marginnote#1{}

\begin{titlepage}
\renewcommand{\thefootnote}{\fnsymbol{footnote}}

\begin{center} \Large
{\bf Theoretical Physics Institute}\\
{\bf University of Minnesota}
\end{center}
\begin{flushright}
TPI-MINN-95/04-T\\
UMN-TH-1330-95\\
hep-ph/9503358\\
March 1995\\
(Revised version)
\end{flushright}
\vspace{.3cm}
\begin{center} \Large
{\bf
Reconciling Supersymmetric
Grand Unification\\ with ${\alpha_s(m_Z)}\approx 0.11$
}
\end{center}

\vspace*{.3cm}

\begin{center} {\Large
L. Roszkowski\footnote{E-mail: {\tt leszek@mnhepw.hep.umn.edu}}
and M.  Shifman\footnote{E-mail: {\tt shifman@vx.cis.umn.edu}}
} \\
\vspace{0.4cm}
{\it  Theoretical Physics Institute, Univ. of Minnesota,
Minneapolis, MN 55455}
\end{center}
\vspace*{1cm}

\begin{abstract}

We argue that supersymmetric grand unification of gauge couplings
is
not incompatible with small ${\alpha_s}$, even without large GUT-scale
corrections, if one relaxes a usual universal gaugino mass
assumption.
A commonly assumed relation ${M_2}\simeq{m_{\widetilde g}}/3$ is in
gross contradiction with ${\alpha_s}\approx0.11$. Instead, small
${\alpha_s}$ favors ${M_2}\gg{m_{\widetilde g}}$. If this is indeed the
case our
observation casts doubt on  another
commonly used relation ${M_1}\simeq 0.5{M_2}$ which originates
from the
same constraint of a common gaugino mass at the GUT scale.
One firm prediction emerging within the small ${\alpha_s}$
scenario with the unconstrained gaugino masses is the existence of a
relatively
light gluino below $\sim$ 200{\rm\,GeV}.

\end{abstract}
\end{titlepage}
\setcounter{footnote}{0}
\setcounter{page}{2}
\setcounter{section}{0}
\newpage

\section{Introduction}

One of the testing grounds for various models of grand unification
is calculating the strong coupling constant ${\alpha_s(m_Z)}$ using, as
input,
the experimental values of the electromagnetic coupling constant
$\alpha$
and ${\sin^2\theta_W}$, where $\theta_W$ is
the Weinberg angle. These calculations have been repeatedly carried
out
in different models and under different assumptions (for  recent
reviews see, {\it e.g.}, Ref.~\cite{Langacker}).
It has been shown, in particular, that the simplest
grand unification
based on the Standard Model (SM) and $SU(5)$ gauge group leads
to too small a value of the strong coupling constant, ${\alpha_s(m_Z)} =
0.073\pm 0.002$~\cite{Langacker2} and
is, thus, ruled out~\cite{F1}.
In contrast, supersymmetric models generally predicted ${\alpha_s(m_Z)}$ in
agreement~\cite{F1} with experimental data available at that time.

A straightforward supersymmetrization of  SM gives rise to the
Minimal Supersymmetric Standard Model (MSSM)~\cite{mssm_review}.
Actually, to fully
specify
the model one has to make an additional assumption about the
pattern of
supersymmetry (SUSY) breaking. The most popular mechanism
is that of soft breaking in which one adds to the Lagrangian
all possible soft SUSY
breaking terms and treats them as independent parameters.
Such terms  arise, {\it e.g.},
when the MSSM is coupled to supergravity~\cite{minisugra}.
This mechanism of generating soft terms is
so deeply rooted that quite often in the current literature
no distinction is made between
the MSSM {\em per se} and the
MSSM plus the assumptions of the minimal
supergravity-based  SUSY breaking. In fact, an overwhelming majority
of papers devoted to even purely
phenomenological studies of the MSSM assume some
(but typically not all) relations stemming from minimal
supergravity, {\it e.g.}, the relation between the mass parameters of the
gauginos of $SU(2)$ and $U(1)$.

Encouraged by early studies~\cite{F1}, many
authors (see, {\it e.g.}, Refs.~\cite{rr,roberts,kkrw1,cmssmstudies})
then studied
unification in the context of the MSSM coupled to minimal
supergravity.
The set of SUSY breaking terms generated this way is
quite restrictive. In particular, in the context of minimal $N=1$
supergravity the masses of all gauginos -- gluinos of $SU(3)$,
winos of $SU(2)$ and the bino of
$U(1)$ --  turn out to be the same at the Planck scale.
Similarly, the soft  mass parameters of all squarks and sleptons
are equal at that scale.
In this restrictive model, which was called
Constrained MSSM (CMSSM)~\cite{kkrw1}, one assumes universal masses
for all the gauginos ($m_{1/2}$) and all the scalars ($m_0$)
at the GUT scale, and often
additionally imposes a mechanism of radiative electroweak
symmetry
breaking (EWSB)~\cite{gsymbreak}.
Accepting these assumptions one arrives at quite definite predictions
for the spectra of masses of the model at the weak scale and for
${\alpha_s(m_Z)}$. For example, the gluino
turns out to be roughly three times heavier than
wino~\cite{mssm_review}. Furthermore,
${\alpha_s(m_Z)}$ generally decreases with increasing $m_{1/2}$ and $m_0$.
Restricting $m_{1/2}$ and $m_0$ (or alternatively all the masses)
below roughly 1{\rm\,TeV} leads to
${\alpha_s(m_Z)}
{\lower.7ex\hbox{$\;\stackrel{\textstyle>}{\sim}\;$}}
0.12$~\cite{kkrw1,Langacker2}.
For example, an updated analysis of Ref.~\cite{lp:new} quotes
${\alpha_s(m_Z)} = 0.129 \pm 0.008$.  The
theoretical error here is
mostly due to  uncertainty associated with the so-called threshold
corrections
at the GUT and low (SUSY breaking)  scales and higher-dimensional
non-renormalizable operators (NRO's)
in the GUT scale Lagrangian.  The
above prediction for  ${\alpha_s(m_Z)}$ was considered as a
great success and the strongest evidence
in favor of the MSSM in  light of
the fact
that, as was believed, the direct measurement of the strong coupling
constant
at LEP and SLD yields ${{\alpha_s(m_Z)}} = 0.125\pm 0.05$~\cite{F2}.

Recently it has been
pointed out, however, that QCD cannot tolerate such
a large
value of the coupling constant~\cite{Shifman}.  A wealth of
low-energy  data
indicates that ${\alpha_s(m_Z)}$ must be very close to
0.11~\cite{Altarelli},
three standard
deviations below the alleged LEP/SLD value.  A method of
determining
${\alpha_s}$ which seems to be  clean theoretically is extracting
${\alpha_s}$ from deep
inelastic scattering (DIS)~\cite{Virchaux}.  A similar
number is obtained in the lattice
QCD~\cite{lattice}. Another reliable
approach
is using~\cite{Eidelman,Voloshin}
(Euclidean) QCD sum rules.  The observation of Ref.~\cite{Shifman}
motivated
a new analysis of  the $\Upsilon$ sum rules~\cite{Voloshin}
claiming the record accuracy achieved so far,
\begin{equation}
{\alpha_s(m_Z)} = 0.109\pm 0.001 \, .
\label{als_voloshin:eq}
\end{equation}
The apparent clash between the low-energy determinations of
the strong coupling constant and those at the $Z$ peak may be
explained~\cite{Shifman} by contributions going beyond SM
which were not taken into account in the global fits.
It should be stressed that the two scenarios --
large ${\alpha_s}$ versus small ${\alpha_s}$ -- cannot coexist
peacefully, as it is sometimes implied in the current literature.
Our starting point is the assumption that the large ${\alpha_s}$
option~\cite{F3},
inconsistent with crucial features of QCD, will eventually
evaporate and the value of the strong coupling
constant at $m_Z$ will stabilize close to 0.11.
In fact, in  Ref.~\cite{Consoli} it has been argued that the
systematic error usually quoted in the LEP number is grossly
underestimated, and that at present LEP experiments can only claim
$0.10{\lower.7ex\hbox{$\;\stackrel{\textstyle<}{\sim}\;$}}
{\alpha_s(m_Z)}{\lower.7ex\hbox{$\;\stackrel{\textstyle<}{\sim}\;$}}0.15$.

The question arises whether grand unification within the framework
of the MSSM can accommodate small ${\alpha_s}\approx0.11$.
This study addresses this question.
Our task is to sort out assumptions
(sometimes implicit) which
are inevitable in analyses of this type
and to find out which assumptions
of the CMSSM absolutely preclude one from
descending to small ${\alpha_s(m_Z)}$ and, therefore, have to
be relaxed.

There are several possible ways to reconcile the prediction for
${\alpha_s(m_Z)}$ in supersymmetric grand unification
with ${\alpha_s(m_Z)}\approx 0.11$.
One is to remain
in the context of the CMSSM but adopt
a heavy SUSY scenario with the SUSY mass spectra significantly
exceeding
1{\rm\,TeV}. This scenario would not only put SUSY into both theoretical
and experimental oblivion
but is also, for the most part, inconsistent
with
our expectations that the lightest supersymmetric particle (LSP)
should be neutral
and/or with the lower bound on the age
of the Universe of at least some 10 billion
years~\cite{kkrw1}. Another possibility is to invoke large enough
negative corrections due to GUT-scale physics.
The issue has been reanalyzed
in a very recent publication~\cite{lp:new}. Under a natural assumption
(the so-called no-conspiracy assumption) it was found that
${\alpha_s(m_Z)}>0.12$. Relaxing this assumption one can, in principle,
construct models of the
CMSSM with large negative contributions coming,
say, from NRO's which could decrease the value of  ${\alpha_s(m_Z)}$
by $\sim 10\%$~\cite{lp:new,urano}. (Alternatively, one can entertain
the possibility of an intermediate scale~\cite{mohapatra} around
$10^{11}{\rm\,GeV}$ whose
existence is motivated by other reasons. In this case, however, many
more unknowns affect the running of the gauge couplings and one cannot
really talk about {\em predicting} ${\alpha_s(m_Z)}$.)
None of these possibilities seem particularly appealing to us.
Although it may well happen that the GUT-scale and NRO corrections
are abnormally large, the guiding idea of grand unification
becomes much less appealing in this case, and the predicitive power
is essentially lost. Indeed,
by appropriately complicating GUT-scale physics one could, perhaps,
achieve gauge coupling unification even in the Standard
Model.

Below we will discuss an alternative route. We will adopt a
down-to-earth, purely phenomenological attitude, with no
assumptions about mechanisms of SUSY breaking.
We do not assume $N=1$ supergravity, nor any mass relations
associated with  this scheme,  for instance, the
equality of the gaugino masses at the GUT scale.
If no theoretical scheme for the mass generation of SUSY partners is
specified one is free to consider any values of these masses.
Our task is to try to find out what pattern of  masses is
 preferred by phenomenology. We consider the MSSM and
limit ourselves to a
``minimal set" of assumptions: (i) all gauge coupling constants are
exactly equal to each other at the GUT scale; (ii) the breaking of
supersymmetry occurs below 1{\rm\,TeV}.

We will show that by
relaxing the CMSSM to the MSSM
one can easily descend to
${\alpha_s(m_Z)} \approx 0.11$. The only effect which is
actually important in dramatically reducing the minimal value of
${\alpha_s(m_Z)}$ is untying the gluino and wino masses.
One firm conclusion is
a relatively light gluino (in the ballpark of 100{\rm\,GeV},
and typically below
200{\rm\,GeV}) and a relatively
heavy wino (at least a few hundred{\rm\,GeV}),
{\it i.e.}, a relation opposite to the one emerging in the CMSSM.
This summarizes our main results.

\section{Calculating ${\alpha_s(m_Z)}$ from grand unification}

\subsection{Procedure}

The procedure for predicting ${\alpha_s(m_Z)}$ assuming gauge coupling
unification
has
been
adequately described in the literature (see, {\it e.g.}, Ref.~\cite{kkrw1}
and references therein), and
we will only summarize it briefly here. The strategy is simple:
the coupling constants $\alpha_1$ and $\alpha_2$ (which are known
more
accurately than ${\alpha_s}$) are evolved from their experimental
values
at $m_Z$ up to the
point where they intersect (which thus defines the unification scale
${M_X}$ and the gauge strength ${\alpha_X}$). At that point one
identifies
${\alpha_s}$ with ${\alpha_X}$ and runs it down to $m_Z$, thus predicting
the value of ${\alpha_s(m_Z)}$ as a function of input parameters.
One-
and two-loop corrections are taken into account.

The renormalization group equations (RGE's) for the gauge couplings
are
given by
\begin{equation}
\frac{d\alpha_i}{d t} = \frac{b_i}{2\pi}\alpha_i^2
+ \mbox{two loops},
\label{rge:eq}
\end{equation}
where $i=1,2,3$,
$t\equiv\log(Q/m_Z)$ and $\alpha_1\equiv\frac{5}{3}\alpha_Y$.
The one-loop
coefficients $b_i$ of the $\beta$ functions for the gauge couplings
change
across each new running mass threshold. In the MSSM they can be
parametrized
as
follows~\cite{rr,gutcorrs,kkrw1}
\begin{eqnarray}
\lefteqn{b_1= \frac{41}{10}
+ \frac{2}{5}\theta_{{\widetilde{H}}}+\frac{1}{10}\theta_{H_2}}
		  \nonumber \\
         & &\mbox{}+\frac{1}{5}\sum_{i=1}^{3}
	\left\lbrace\frac{1}{12}\left(
	        \theta_{{\tilde{u}}_{L_i}}
		+ \theta_{{\tilde{d}}_{L_i}}\right)
		+ \frac{4}{3}\theta_{{\tilde{u}}_{R_i}}
		+ \frac{1}{3}\theta_{{\tilde{d}}_{R_i}}
		+ \frac{1}{4}\left(\theta_{{\tilde{e}}_{L_i}}
		+ \theta_{{\tilde{\nu}}_{L_i}}\right)
		+ \theta_{{\tilde{e}}_{R_i}}\right\rbrace
		\label{b1:eq} \\
\lefteqn{b_2= -\frac{19}{6}
		+ \frac{4}{3}\theta_{\widetilde W}
		+ \frac{2}{3}\theta_{\widetilde{H}} + \frac{1}{6}\theta_{H_2}
+ \frac{1}{2}\sum_{i=1}^3\left\lbrace\theta_{{\tilde{u}}_{L_i}}
	          \theta_{{\tilde{d}}_{L_i}}
		+ \frac{1}{3}\theta_{{\tilde{e}}_{L_i}}
		  \theta_{{\tilde{\nu}}_{L_i}}\right\rbrace }
		\label{b2:eq} \\
\lefteqn{b_3= -7 + 2\,\theta_{\widetilde g}
	+ \frac{1}{6}\sum_{i=1}^3\left\lbrace\theta_{{\tilde{u}}_{L_i}}
		+ \theta_{{\tilde{d}}_{L_i}}
		+ \theta_{{\tilde{u}}_{R_i}} +
\theta_{{\tilde{d}}_{R_i}}\right\rbrace}
		\label{b3:eq}
\end{eqnarray}
where
$\theta_x\equiv\theta(Q^2-m_x^2)$.

In Eqs.~(\ref{b1:eq})--(\ref{b3:eq}) ${\tilde{H}}$ stands for the (mass
degenerate) higgsino fields,
$\widetilde{W}$ for the winos, the partners of the $SU(2)$ gauge bosons
($m_{\widetilde{W}}\equiv{M_2}$),
and ${\widetilde g}$ stands for the gluino, all taken to be mass eigenstates
in this
approximation.
Also, in this approximation $H_2$ stands for
a heavy Higgs
doublet, as explained in Ref.~\cite{kkrw1}.
(The full 2-loop gauge
coupling $\beta$-functions for the SM and the MSSM which we use in
actual
calculations can be found, {\it e.g.},  in
Ref.~\cite{bbo}.)

Eqs.~(\ref{b1:eq})--(\ref{b3:eq}) represent so-called leading log
approximation and involves some simplifications.
However, as we will argue later,
it will be
sufficient to present the basic points of our analysis and answer the
question how low  one could descend in the  values of ${\alpha_s(m_Z)}$
assuming only
strict unification of the  gauge couplings in  the MSSM.

The prediction for ${\alpha_s(m_Z)}$ depends on the adopted values of the
input parameters: $\alpha$, ${\sin^2\theta_W}(m_Z)$, and ${m_t}$. It also
receives
corrections from: the two-loop gauge and Yukawa contributions,
scheme
dependence (${\rm\overline{MS}}$ {\it versus} ${\rm\overline{DR}}$), mass
thresholds at the
electroweak scale and,
finally, the  GUT-scale mass thresholds and NRO contributions.
 We will
discuss these effects in turn now.

The input values of $\alpha_1$ and $\alpha_2$ at
$Q=m_Z$ can be extracted from the experimental values of
$\alpha(m_Z)$
and ${\sin^2\theta_W}(m_Z)$.
For the electromagnetic coupling we
take~\cite{pdb}
\begin{equation}
\alpha(m_Z)={1\over{127.9\pm0.1}}.
\label{alphaeminput:eq}
\end{equation}
Recently, three groups have reanalyzed
$\alpha(m_Z)$~\cite{alpha_recent}
and obtained basically similar results: $\alpha(m_Z)^{-1}=
127.96\pm0.06$
(Martin and Zeppenfeld),
$127.87\pm0.10$
(Eidelman and Jegerlehner), and $128.05\pm0.10$ (Swartz).
Adopting even
the largest (central) value of Swartz would shift ${\alpha_s(m_Z)}$
up by only
0.001~\cite{lp:new}.

The range of input values of ${\sin^2\theta_W}(m_Z)$ is rather
critical. This sensitivity is due to the fact that  $\alpha_2(Q)$ does
not change
between
$Q=m_Z$ and the GUT scale $Q={M_X}$ as much as the other two
couplings.
Thus, a small increase in ${\sin^2\theta_W}(m_Z)$ has an enhanced (and
negative)
effect on
the resulting value of ${\alpha_s(m_Z)}$.
Following Ref.~\cite{lp:new}
we assume~\cite{F4}
\begin{equation}
{\sin^2\theta_W}(m_Z)=0.2316\pm0.0003 - 0.88\times10^{-7}{{\rm\,GeV}}^2
\left[{m_t}^{2}
- (160{\rm\,GeV})^{2} \right].
\label{s2winput:eq}
\end{equation}
Moreover, the  global analysis
of  Ref.~\cite{EL} implies that in the MSSM
${m_t}=160\pm13{\rm\,GeV}$.
Recently, both the CDF and D0 collaborations have reported
discovery of the top
quark
and quoted somewhat higher mass ranges:
${m_t}=176 \pm8\pm10{\rm\,GeV}$ (CDF)~\cite{cdf:top} and ${m_t}= 199\pm
20\pm22{\rm\,GeV}$
(D0)~\cite{dzero:top}.
Such high (central)
values of ${m_t}$ would lower ${\sin^2\theta_W}(m_Z)$ and {\em increase}
${\alpha_s(m_Z)}$
by 0.002 and 0.005, respectively.

Including the two-loop terms in the RGE's increases ${\alpha_s(m_Z)}$ by
about
10\%.
There are two types of contributions to ${\alpha_s(m_Z)}$ at the two-loop
level. Pure gauge term yields $\Delta{\alpha_s(m_Z)}=0.012$ if one
assumes
SUSY in both one-  and two-loop  coefficients of the $\beta$ function
all the way down
to
$Q=m_Z$. This is the most important correction to the one-loop value
of
${\alpha_s(m_Z)}$.
If, instead, the two-loop coefficients of the pure gauge part are
changed
to their SM  values at $Q=1{\rm\,TeV}$, one finds an additional shift
$\Delta{\alpha_s(m_Z)}\approx 0.0007$. Since this shift is negligibly small,
we  keep the two-loop  coefficients supersymmetric all the way
down to
$m_Z$.
Corrections due to the Yukawa-coupling contribution to the RGE's are
also small, although negative~\cite{lp:new}.
In the limit of large top Yukawa coupling
(${h_t}\simeq1$, ${h_b}\simeq0\simeq {h_\tau}$, as in the small
$\tan\beta\simeq1$
scenario)
one finds $\Delta{\alpha_s(m_Z)}=-0.0015$ while even in the extreme case
of
the large $\tan\beta$ scenario
(${h_t}\simeq {h_b} \simeq {h_\tau}\simeq1$)
$\Delta{\alpha_s(m_Z)}=-0.004$, in agreement with
Ref.~\cite{lp:new}.

Above $Q=1${\rm\,TeV} we also change from the conventional ${\rm\overline{MS}}$
scheme,
that we use throughout this paper, to the fully supersymmetric
${\rm\overline{DR}}$ scheme. The corresponding shift in ${\alpha_s(m_Z)}$ is
about
0.0002
and is negligible numerically~\cite{Langacker2,kkrw1,gutcorrs}.

Before proceeding to discussing in more detail the contribution from
one-loop threshold effects,  a remark is in order on possible
corrections from the GUT-scale mass thresholds and NRO's. Since in
this
paper we look for an alternative way of lowering ${\alpha_s(m_Z)}$, we
switch off  all corrections from the GUT-scale physics whatsoever. As
was noted
previously~\cite{gutcorrs,Langacker2,urano,lp:new} they are
GUT-model dependent and, in principle, can be
sizeable.  For instance, according to Refs.~\cite{Langacker2,lp:new}
the corresponding effect in  ${\alpha_s(m_Z)}$ can be as large as
$\sim0.008$;
a factor of 2.5 larger effect is needed, however, to ensure
${\alpha_s(m_Z)}\approx 0.11$.
Building a fully elaborated
and phenomenologically acceptable  model of this type seems to be a
task for the
future.

What remains to be done is to explain our treatment of the mass
thresholds at the electroweak and SUSY scales.  We use usual
the step-like
approximation in the coefficients of the
$\beta$-function, Eqs.~(\ref{b1:eq})--(\ref{b3:eq}).
In the one-loop coefficients the jumps occur
at
the positions of the masses of the individual particles while in the
two-loop coefficients it is sufficient, to our accuracy, to consider
one jump at a common SUSY scale, as explained above.
As a matter of fact,
with no loss of accuracy, we take this scale in the two-loop
coefficients to be lower than $m_Z$ so that in our evolution from
${M_X}$ down to $m_Z$ we treat the two-loop coefficients as fully
supersymmetric. Also, the $t$ quark is not frozen at ${m_t}$ in the
two-loop coefficients. It is well
known that the step-like approximation is not absolutely accurate in
the problem of the coupling constant evolution
(see, {\it e.g.}, Ref.~\cite{kataev}
for a recent discussion), especially if the mass thresholds are
rather close to $m_Z$, as is the case with $t$ quark.
We find that the other thresholds
are far less important,
since,  as we vary their positions, the effect of
the variation mimics the non-logarithmic corrections omitted in the
step approximation. The error in ${\alpha_s(m_Z)}$ due to the
inacuracy of our approximation of the ${\alpha_s}$ evolution   at
${m_t}$ is less than 1\% and is, thus, unimportant.

\subsection{MSSM with gauge unification only}

The question we want to address is whether supersymmetric grand
unification
necessarily predicts large values of
${\alpha_s(m_Z)}{\lower.7ex\hbox{$\;\stackrel{\textstyle>}{\sim}\;$}}0.12$
as long as
all SUSY masses are restricted to lie below 1{\rm\,TeV}. This is indeed
the case in the
CMSSM with additional assumptions of common gaugino mass and
common
scalar mass, as described in the Introduction.

In order to track the role of these mass relations we begin by
treating the masses of
the different types of
states as completely independent parameters.
We choose to remain open-minded and not biased
by
any additional (even  well-motivated)
assumptions about  the parameters involved, other
than the basic idea of gauge coupling unification.
Thus, we
assume no relation between squarks and sleptons, or between the
gauginos.
(Actually, the structure of supersymmetry alone forces certain
relations between sfermion masses and gaugino masses, thus
disallowing, for example, very light squarks and very
heavy gauginos~\cite{ibanez}.
We will see below that this will not have any substantial effect on
our results.)
We also do not impose a mechanism of radiative electroweak
symmetry
breaking. We will see {\it a posteriori} that requiring EWSB
will not change our conclusions significantly.

\begin{figure}
\centering
\epsfxsize=6in
\hspace*{0in}
\epsffile{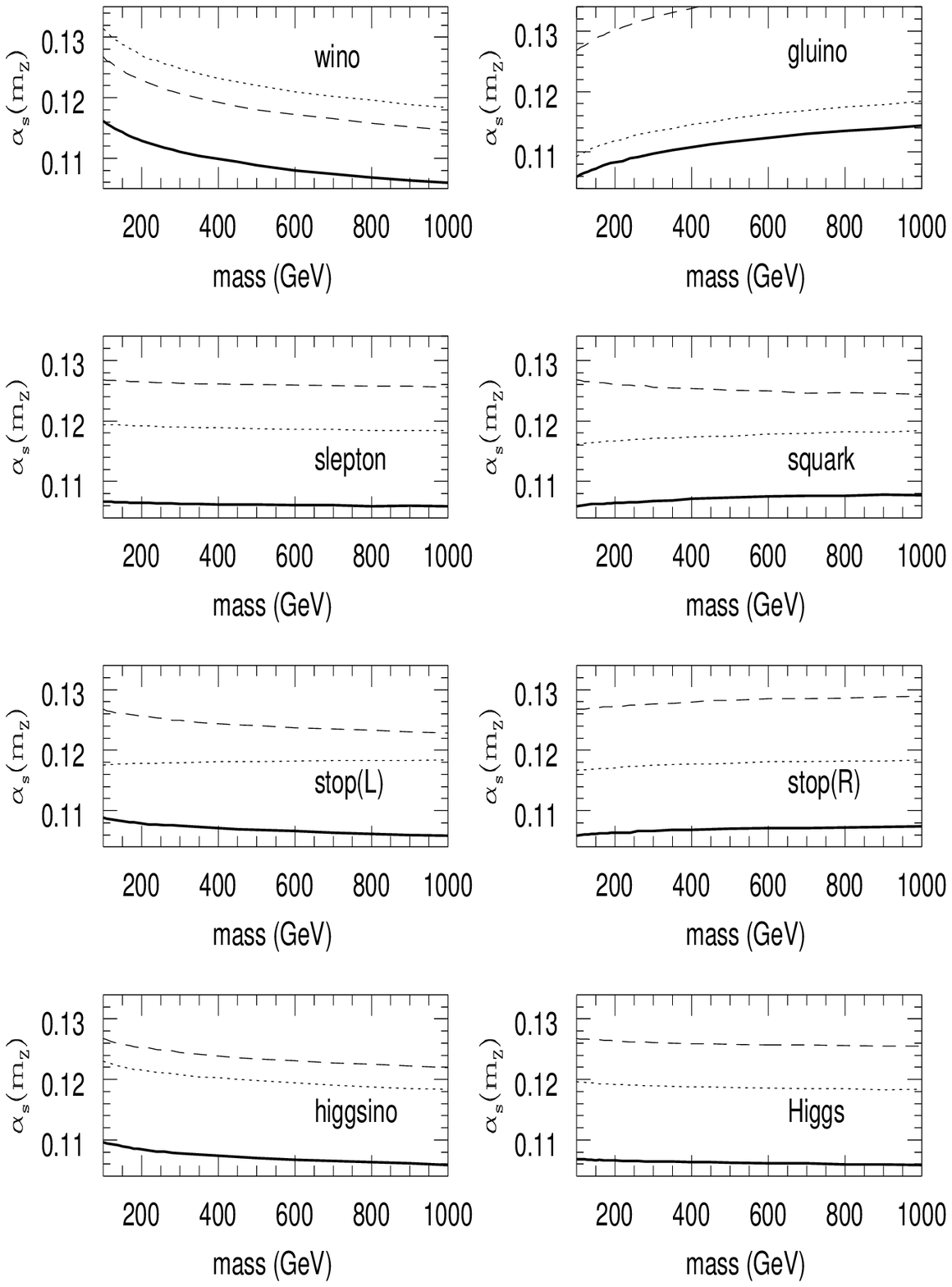}
\caption{Dependence of ${\alpha_s(m_Z)}$ on the mass of individual states
entering the one-loop thresholds, as in
Eqs.~(\protect{\ref{b1:eq}})--(\protect{\ref{b3:eq}}). The masses of
all other states are set to either 100{\rm\,GeV} (dash) or 1{\rm\,TeV} (dots)
and
${m_t}=160{\rm\,GeV}$. Also plotted (thick solid) is
${\alpha_s^{\rm min}(m_Z)}$ -
the lowest range of ${\alpha_s(m_Z)}$ obtained by choosing other mass
parameters in such a way as to minimize it (as in the last row of
Table~\protect{\ref{als:table}}).
}
\label{allmass:fig}
\end{figure}

In Fig.~\ref{allmass:fig} we show ${\alpha_s(m_Z)}$ as a function of the
mass of each relevant type of state. We assume all other masses to
be
degenerate and equal to either 100{\rm\,GeV} or 1{\rm\,TeV}.
Generally, we will treat all squarks and all sleptons
as mass-degenerate.
The only exception to this rule will be the scalar top states,
${{\tilde t}_L}$ and ${{\tilde t}_R}$. This is because their masses are
typically
expected to be
significantly different from the other squarks and from each other.

\begin{table}
\centering
{\footnotesize 
\begin{tabular}{|c|c|c|c|c|c|c|c||c|}
\hline
${M_2}$ & ${m_{\widetilde g}}$ & ${m_{\tilde{l}}}$ & ${m_{\tilde{q}}}$ &
${m_{{\tilde t}_L}}$ & ${m_{{\tilde t}_R}}$ & $m_{\widetilde H}$ & $m_{H_2}$
& ${\alpha_s(m_Z)}$ \\
\hline\hline
100{\rm\,GeV} & 100{\rm\,GeV} & 100{\rm\,GeV} & 100{\rm\,GeV} & 100{\rm\,GeV}
& 100{\rm\,GeV} &
100{\rm\,GeV} &  100{\rm\,GeV} & 0.127\\ \hline
500{\rm\,GeV} & 100{\rm\,GeV} & 100{\rm\,GeV} & 100{\rm\,GeV} & 100{\rm\,GeV}
& 100{\rm\,GeV} & 100{\rm\,GeV} &  100{\rm\,GeV}
&
0.118\\ \hline
1{\rm\,TeV} & 1{\rm\,TeV} & 1{\rm\,TeV} & 1{\rm\,TeV} & 1{\rm\,TeV}
& 1{\rm\,TeV} & 1{\rm\,TeV} & 1{\rm\,TeV} &
0.118 \\ \hline
1{\rm\,TeV} & 500{\rm\,GeV} & 1{\rm\,TeV} & 100{\rm\,GeV} & 1{\rm\,TeV}
& 100{\rm\,GeV} & 1{\rm\,TeV} & 1{\rm\,TeV} &
0.112\\ \hline
\hline
1{\rm\,TeV} & 100{\rm\,GeV} & 1{\rm\,TeV} & 100{\rm\,GeV} & 1{\rm\,TeV}
& 100{\rm\,GeV} & 1{\rm\,TeV} & 1{\rm\,TeV} &
0.106\\ \hline\hline
\end{tabular}}
\caption{${\alpha_s(m_Z)}$ for several choices of mass parameters (assumed
between 100{\rm\,GeV} and 1{\rm\,TeV}) and
${m_t}=160{\rm\,GeV}$. Last row displays the case for which the smallest
${\alpha_s(m_Z)}$ was found.
}
\label{als:table}
\end{table}

It is obvious from the form of the $\beta$-functions,
Eqs.~(\ref{b1:eq})--(\ref{b3:eq}), that the resulting value of
${\alpha_s}(m_Z)$ will most sensitively depend on two parameters
only:
the gluino
mass ${m_{\widetilde g}}$ and the soft mass parameter ${M_2}$ of the wino.
The
reasons are twofold: not only are their $\beta$-function coefficients
among the largest but also they change only one out of the three
$b_i$'s.
Fig.~\ref{allmass:fig} clearly confirms our expectation.
Also, Table~\ref{als:table} shows ${\alpha_s(m_Z)}$ for several
choices of
relevant parameters. The first four rows are meant to demonstrate
the dependence of ${\alpha_s(m_Z)}$ on
${M_2}$ and ${m_{\widetilde g}}$.

We are interested in the lowest possible values of ${\alpha_s(m_Z)}$
allowed by (strict) grand unification. As it  is obvious
from  Fig.~\ref{allmass:fig}, minimization of
${\alpha_s(m_Z)}$ requires  minimizing
${m_{\widetilde g}}$ and
${m_{{\tilde t}_R}}$ while simultaneously maximizing
the masses
of the wino, the sleptons, the higgsino, and of the heavy Higgs.
We have also verified that, in order to
minimize
${\alpha_s(m_Z)}$, one should also set ${m_{\tilde{q}}}$
(${m_{{\tilde t}_L}}$) at its
lowest (largest)
possible value. Since the ``standard"  prediction for ${\alpha_s(m_Z)}$
emerging in the
CMSSM is quoted above under the assumption that all sparticles are
lighter than 1{\rm\,TeV}
we accordingly restrict all the masses to that
range.
At the lower end, we allow the masses to lie as low as $100{\rm\,GeV}$.
(Lowering
this
limit down to $m_Z$ would not noticeably change
${\alpha_s(m_Z)}$~\cite{bagger}.)
In the last row of Table~\ref{als:table}
we show the lowest value of
${\alpha_s(m_Z)}$ obtained by varying all the mass parameters between
100{\rm\,GeV} and 1{\rm\,TeV}.
Experimental bounds on most of those states are still  less than
$m_Z$. Even for ${m_{\widetilde g}}$ and the masses of the squarks there are
no
inescapable lower bounds, other than roughly $m_Z/2$ from
LEP~\cite{galtieri}.
(Very recently, the D0 collaboration~\cite{galtieri} has published new
improved limits: ${m_{\widetilde g}}>144{\rm\,GeV}$ for any
${m_{\tilde q}}$
and ${m_{\widetilde g}}>212{\rm\,GeV}$ for ${m_{\widetilde g}}={m_{\tilde
q}}$. Adopting these limits in the last row of Table~\ref{als:table}
would increase ${\alpha_s^{\rm min}(m_Z)}$ by only 0.002 and 0.003,
respectively.)

We also display in
Fig.~\ref{allmass:fig} ${\alpha_s^{\rm min}(m_Z)}$ (thick solid line)
as a function of the mass of
each individual
state, while setting all the other masses as
in the last row of Table~\ref{als:table}.
It is clear that in general one can easily
obtain values of ${\alpha_s(m_Z)}$ small enough to accomodate the
range ${\alpha_s(m_Z)}\approx0.11$ which  we favor. Furthermore,
${\alpha_s(m_Z)}$ shows little dependence on the masses of the states
other
than the $SU(2)$ and $SU(3)$ gauginos. Therefore
one actually has considerable
freedom in choosing the other masses as desired. This justifies our
approach of assuming all sleptons to be mass-degenerate, and
similarly
with squarks. Furthermore, relatively weak dependence of
${\alpha_s(m_Z)}$
on the mass of the higgsino (which we approximate by the
Higgs/higgsino mass parameter $\mu$) shows that imposing EWSB
would
probably not lead to any strong increase in
the lower bound on ${\alpha_s(m_Z)}$. This is
because the conditions of EWSB determine $\mu$ in terms of (soft)
Higgs
mass parameters which influence ${\alpha_s(m_Z)}$ even less.

It is also
evident from the gluino window of Fig.~\ref{allmass:fig}
that the mass of the gluino is
strongly confined to rather small values in the range of a few
hundred {\rm\,GeV} only. This is a distinctive feature and a strong
prediction of our approach. The exact value of the upper bound on
${m_{\widetilde g}}$ that one allows clearly depends on how large GUT-
related
corrections one assumes and also how large values of
${\alpha_s(m_Z)}$ one is willing to accept.

\begin{figure}
\centering
\epsfxsize=3.5in
\hspace*{0in}
\epsffile{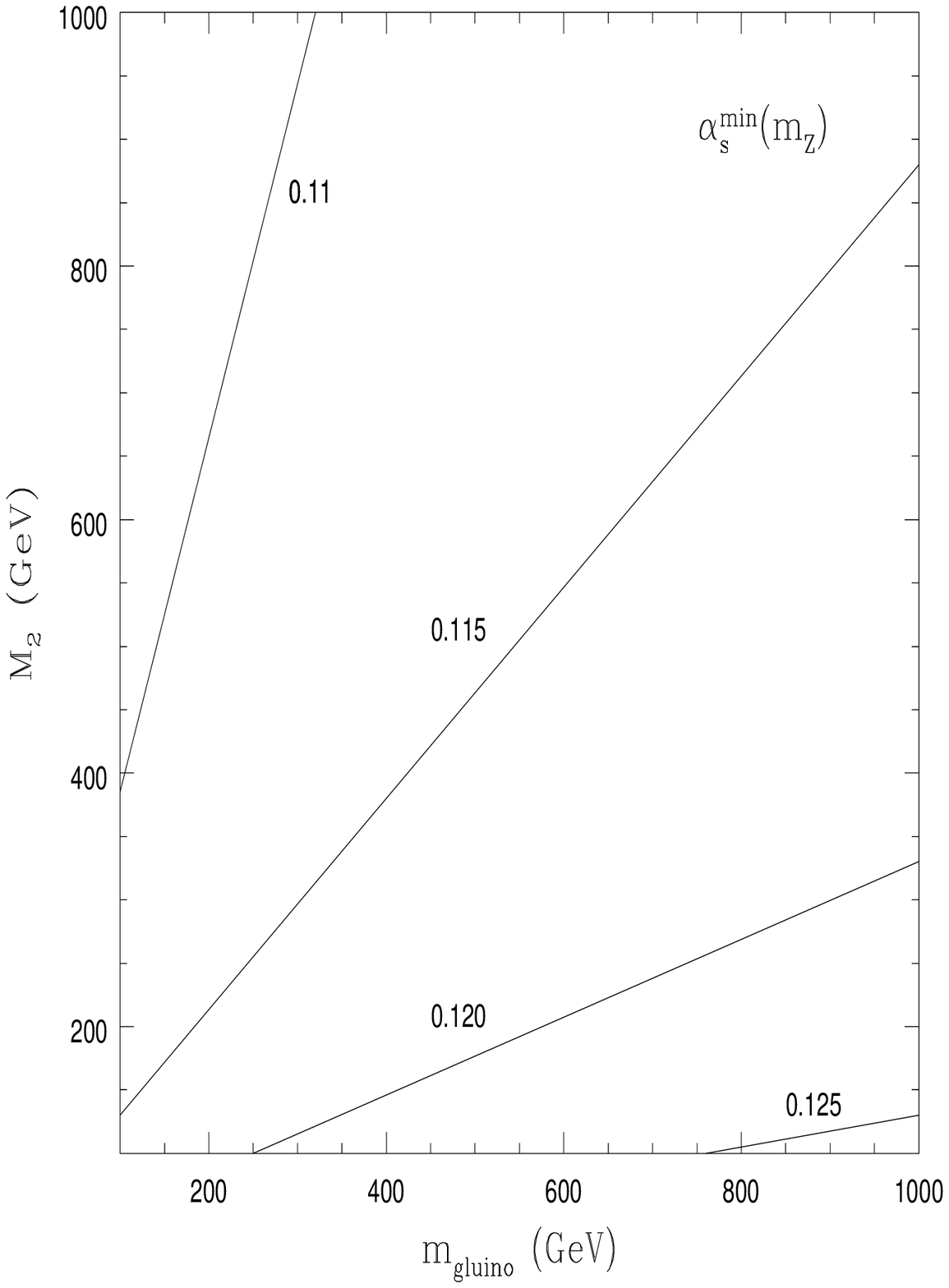}
\caption{Contours of constant ${\alpha_s^{\rm min}(m_Z)}$ in the
(${m_{\widetilde g}},{M_2}$)
plane. All other mass parameters are chosen so as to minimize
${\alpha_s(m_Z)}$ (as in the last row of
Table~\protect{\ref{als:table}}) and  ${m_t}=160{\rm\,GeV}$.
}
\label{winogluino:fig}
\end{figure}

On the other hand, the wino mass parameter ${M_2}$ should
preferably
be larger than ${m_{\widetilde g}}$, contrary to what is commonly expected.
This is clearly shown in
Fig.~\ref{winogluino:fig} where, in the plane
(${m_{\widetilde g}},{M_2}$), we plot the lowest allowed values of
${\alpha_s(m_Z)}$
found by assuming all other mass parameters as in the last
row of Table~\ref{als:table}.
It is clear that ${\alpha_s(m_Z)}\approx0.11$ favors
relatively small ${m_{\widetilde g}}$ and large ${M_2}$.

\subsection{Relating gaugino masses}

Among perhaps the
most commonly assumed, and least questioned, relations
are
the
ones between the mass parameters of the gauginos
\begin{eqnarray}
{M_1}&=& {5\over3}\tan^2\theta_{\rm W}\,{M_2}\simeq\,0.5 {M_2},
\label{monemtwo:eq} \\
{M_2} &=& \frac{\alpha_2}{{\alpha_s}}{m_{\widetilde g}}
\simeq\,0.3{m_{\widetilde g}},
\label{mtwomgluino:eq}
\end{eqnarray}
where the SUSY breaking parameters
${M_1}$, ${M_2}$ and ${m_{\widetilde g}}$ of
the bino, the wino, and the gluino states
are evaluated at the electroweak
scale.
Virtually all phenomenological and experimental studies adopt at
least
the relation~(\ref{monemtwo:eq}).
Strictly speaking, however, both relations are not necessary in the
context of the MSSM.
They
both originate from the assumption that, in minimal $SU(5)$ $N=1$
supergravity, the kinetic
term of the gauge bosons and gauginos is equal to a Kronecker delta.
Clearly, {\it a priori} this assumption is not an indispensible  part
of the MSSM.

\begin{figure}
\centering
\epsfxsize=3.5in
\hspace*{0in}
\epsffile{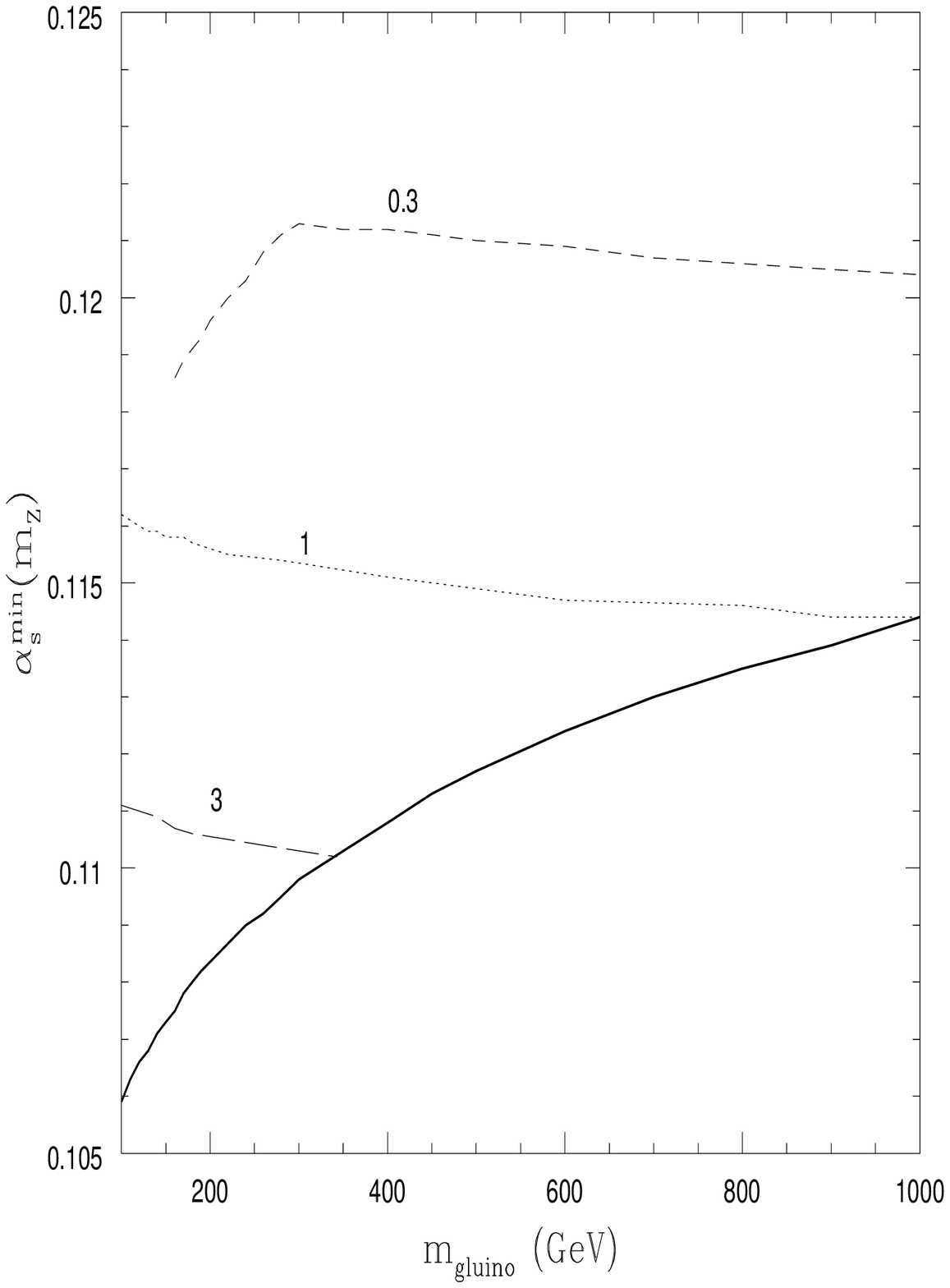}
\caption{${\alpha_s^{\rm min}(m_Z)}$ versus ${m_{\widetilde g}}$ for several
choices
of $x$ assuming ${M_2}=x\,{m_{\widetilde g}}$. All other mass
parameters are set in such a way  as to minimize ${\alpha_s(m_Z)}$
(as in the last row of
Table~\protect{\ref{als:table}}), except ${m_t}=160{\rm\,GeV}$.
For $x=0.3$, the range ${m_{\widetilde
g}}{\lower.7ex\hbox{$\;\stackrel{\textstyle<}{\sim}\;$}}157{\rm\,GeV}$
corresponds to
(wino-like) chargino lighter than about 47{\rm\,GeV} excluded by LEP.
For $x=3$, ${m_{\widetilde
g}}{\lower.7ex\hbox{$\;\stackrel{\textstyle<}{\sim}\;$}}333{\rm\,GeV}$ from
requiring ${M_2}<1{\rm\,TeV}$.
As in Fig.~\protect{\ref{allmass:fig}}
the thick solid curve represents ${\alpha_s^{\rm min}(m_Z)}$ -
the lowest range of ${\alpha_s(m_Z)}$ obtained by choosing mass
parameters, other than ${m_{\widetilde g}}$,
in such a way as to minimize it (as in the last row of
Table~\protect{\ref{als:table}}).
The
value
$x\approx 0.3$ represents the choice commonly made in the
literature.
}
\label{xgluinogut:fig}
\end{figure}

{}From our previous analysis it is evident that any additional
assumption
relating the masses of the wino and the gluino will have a significant
impact on the prediction of ${\alpha_s(m_Z)}$. In Fig.~\ref{xgluinogut:fig}
we plot ${\alpha_s^{\rm min}(m_Z)}$ versus ${m_{\widetilde g}}$ for
${M_2}=x\,{m_{\widetilde g}}$. We set all the other masses in such a
way as to
minimize
${\alpha_s(m_Z)}$, as in the last row of Table~\ref{als:table}.
We also show the lowest allowed
${\alpha_s(m_Z)}$
(thick solid curve) as a function of ${m_{\widetilde g}}$ only by setting also
${M_2}=1{\rm\,TeV}$.
It is clear that the usually assumed ratio $x\approx 0.3$
forces
${\alpha_s(m_Z)}$ above $\sim0.120$. To be consistent with
${\alpha_s(m_Z)}\approx 0.11$ the ratio $x{\lower.7ex\hbox{$\;\stackrel
{\textstyle>}{\sim}\;$}}3$ is required.
This corresponds to
${M_2}{\lower.7ex\hbox{$\;\stackrel{\textstyle>}{\sim}\;$}}
9\,{m_{\widetilde g}}$ at the GUT scale.

Furthermore, Fig.~\ref{xgluinogut:fig} shows that
the mass of the gluino must again be rather small,
${m_{\widetilde g}}{\lower.7ex\hbox{$\;\stackrel{\textstyle<}{\sim}\;$}}
300{\rm\,GeV}$, in the absence of large GUT-scale
corrections,
unless one allows for
the wino mass parameter ${M_2}$ significantly above
1{\rm\,TeV}.

The above considerations put into doubt also the
relation~(\ref{monemtwo:eq}), which has its root in the same
assumption of the equality of all the gaugino masses at the GUT
scale.
It is true that the mass parameter of the bino ${M_1}$ does not
enter
Eqs.~(\ref{b1:eq})--(\ref{b3:eq}) and cannot be directly related to
${M_2}$ and ${m_{\widetilde g}}$.
However, in
the CMSSM the lightest neutralino almost
invariably comes out to be an almost pure bino~\cite{roberts,kkrw1}
and
${m_\chi}\simeq{M_1}$. It is also an excellent dark matter
candidate. There are also stringent limits on the cosmic abundance of
exotic particles with color and electric charges.
Requiring
that the lightest (bino-like) neutralino be lighter than the gluino,
and thus a likely candidate for the lightest supersymmetric
particle (LSP) leads to ${M_1}{\lower.7ex\hbox{$\;\stackrel{\textstyle<}{\sim}
\;$}}{1\over3}{M_2}$
(or ${M_1}{\lower.7ex\hbox{$\;\stackrel{\textstyle<}{\sim}\;$}}
{2\over3}{M_2}$ at ${M_X}$), thus violating
the
relation~(\ref{monemtwo:eq})~\cite{dennis}.

Many phenomenological and dark matter properties of the neutralinos
depend on the relation~(\ref{monemtwo:eq}). Relaxing it may bear
important consequences for neutralino detection in
accelerators~\cite{gr,majerotto}  and
in dark matter searches~\cite{gr}, as well as in placing bounds on
other sparticles. Basically, the mass of
the (lightest) bino-like neutralino is ${m_\chi}\simeq{M_1}$. Reducing the
ratio
${M_1}/{M_2}$ leads to lighter neutralinos. The region of the
plane ($\mu,{M_2}$) (as it is usually presented) where $\chi$ remains
mostly bino-like actually increases somewhat~\cite{gr}. Also, even
rather light neutralinos with mass in the range 3{\rm\,GeV} to a few tens
of {\rm\,GeV} are in principle not excluded and possess excellent dark
matter properties (${\Omega_\chi h_0^2}\sim1$)~\cite{gr}.

Finally, it is worth commenting that, even in the context of $N=1$
supergravity one can  relax the
assumptions~(\ref{monemtwo:eq})--(\ref{mtwomgluino:eq})~\cite{eent,drees}.
This can be done by considering a general form of the kinetic term of
the gauge and gaugino fields, rather than assuming it to be equal to
unity.
In this case one finds that the gauge couplings at
${M_X}$ need not be equal (thus making the GUT energy scale ${M_X}$
somewhat ill-defined) and, in general, relations among
gaugino masses become arbitrary. If, however, one assumes ${M_X}\ll m_{\rm
Planck}$ then one finds, at ${M_X}$, ${m_{\widetilde g}}/{\alpha_s}=
-\frac{3}{2}{M_2}/\alpha_2 + \frac{5}{2}{M_1}/\alpha_1$~\cite{eent}.
In the limit in which the gauge couplings are only slightly displaced
from each other at ${M_X}$ we find
$({m_{\widetilde g}}/{M_2})_{|_{M_X}}\simeq-\frac{3}{2}
+ \frac{5}{2}({M_1}/{M_2})_{|_{M_X}}$.
One solution is the usual ${m_{\widetilde g}}={M_2}={M_1}$. But there
exist also solutions to this relation which are consistent with small
${\alpha_s(m_Z)}$, for example $({m_{\widetilde g}}/{M_2})_{|_{M_X}}\simeq0.1$
and
$({M_1}/{M_2})_{|_{M_X}}\simeq0.64$, in agreement with what we have
found above. Thus it may be possible
to reconcile ${\alpha_s(m_Z)}\approx0.11$ with some non-minimal
versions on $N=1$ supergravity.

\section{Phenomenological consequences}

The version of supersymmetric grand unification considered here
leads
to several distinct implications. One is the necessary existence of a
relatively light gluino below $\sim$ 200 {\rm\,GeV} and
preferably large wino
mass parameter ${M_2}$. The likely
violation of the commonly assumed
relations~(\ref{monemtwo:eq})--(\ref{mtwomgluino:eq}) may lead to many
important consequences for placing bounds on various sparticles
and to more promissing prospects for neutralino dark matter searches.

Below we discuss how the existence of a light gluino affects possible
solutions to the long-lasting anomaly of the $Z\rightarrow b\bar b $
width.
Furthermore, ${\alpha_s}\approx 0.11$ may lead to
a significant relaxation of the constraints on $\tan\beta$ from requiring
$b$--$\tau$ mass unification.
We discuss these points below.

\subsection{Consequences of light gluino}

If ${\alpha_s(m_Z)}\approx 0.11$ does indeed require the gluino mass
to lie in the ballpark of 100{\rm\,GeV}, as was argued above,  the question
which immediately comes to one's mind is: ``what are other
phenomenological implications of such a light gluino?"

First and foremost, with this mass, the gluino must be accessible to
direct
searches at the Tevatron. Currently, a gluino mass range up to about
200{\rm\,GeV}
is probed~\cite{galtieri} but no firm assumption-independent bounds
can be drawn. On the other hand, with the Main Injector upgrade, the
Tevatron experiments will be able to probe ${m_{\widetilde g}}$ in the range up
to 300{\rm\,GeV}. If the gluino is indeed found below some 240{\rm\,GeV} and no
(wino-like) chargino is found at LEP-II up to some 80{\rm\,GeV}, we will
know that the relation~(\ref{mtwomgluino:eq}) does not hold.

Second, light gluinos propagating in loops make the corresponding
radiative
corrections more pronounced. They can then
become important in understanding several facts
where hints on disagreement between observations and SM
expectations were  detected.  The most well-known example of this
type is the problem of ${\alpha_s}$ itself.  As was noted in
Refs.~\cite{Hagi,Djou} the gluino exchange correction to the
$Zq\bar q$ vertices is positive so that the gluino correction enhances
the hadronic width of $Z$, imitating in this way a larger value of
${\alpha_s}$. Fig.~2 of Ref.~\cite{Hagi} shows
that the correction can reach $\sim 0.4\%$ in each quark channel
provided that ${m_{\widetilde g}}\sim 100{\rm\,GeV}$ and ${m_{\tilde{q}}}
\sim70{\rm\,GeV}$.
With such a  correction the value of ${\alpha_s}$ measured
at the $Z$ peak slides down by $\sim 10\%$ solving the problem in
full.

On the other hand, it seems extremely unlikely that the very same
mechanism may be responsible for the alleged enhancement in the
$b\bar b$ channel. Indeed, if we take the central value for
the experimental $Z\rightarrow b\bar b $ width, the excess over the
theoretical expectation amounts to $\sim 7$ MeV~\cite{Langacker}, a
factor of 5 larger
than the excess produced by the gluino correction above. One would
have to descend to unacceptably low squark and gluino masses to get
this factor of 5. Recently,
 another possible solution of  the $R_b$ problem was suggested in
Ref.~\cite{kkw}.  In this work
the mass parameters of  the
MSSM were also considered as {\it a priori}
unrelated. It was
shown that, in order to induce large enough SUSY correction to
reconcile the measured value of $R_b$ with the SM prediction, a
relatively
light (below roughly 80{\rm\,GeV}) higgsino-like chargino is required. The
authors
also need at least one stop with a significant ${{\tilde t}_R}$ component in
the same mass range. In order to examine what predictions for
${\alpha_s(m_Z)}$ this scenario leads to we have set the higgsino mass
parameter $\mu$ and ${m_{{\tilde t}_R}}$ at $m_Z$, and chosen all other mass
parameters in such a way as to minimize ${\alpha_s(m_Z)}$, as before. We
find
${\alpha_s(m_Z)}{\lower.7ex\hbox{$\;\stackrel{\textstyle>}{\sim}\;$}}0.11$.

Another problem where the relatively light gluino can help is the
deficit of
the semileptonic branching ratio in $B$ mesons and the charm
multiplicity~\cite{BBSV}.  Theoretical calculations of these
quantities
are at a rather advanced stage now. Both perturbative
and non-perturbative effects have been considered. The most detailed
analysis of the
non-perturbative effects is carried out in Ref.~\cite{BBSV},
with the conclusion that they can be essentially neglected in the
problem at hand. As for perturbative calculations, they have been
repeatedly discussed in the literature. (See, {\it e.g.}, recent
papers~\cite{Ball,Ball2} and references therein.)
The theoretical prediction
turns out to be rather sensitive to the choice of the value of
${\alpha_s}$ and the normalization scale $\mu$ relevant to the
process. Smaller values of $\mu$ and larger values ${\alpha_s}$
tend to enhance the non-leptonic width and, thus, lower the
prediction for the semileptonic branching ratio.  On the contrary,
larger values of $\mu$ and smaller ${\alpha_s}$ suppress the
non-leptonic width and enhance the branching ratio.  The theoretical
prediction can be made marginally compatible~\cite{Ball2} with the
data
 on the semileptonic branching ratio~\cite{glasgow} provided that
${\alpha_s}$ is chosen on the high side and $\mu$ on the low side. At
the same time, if ${\alpha_s(m_Z)}\approx 0.11$ the prediction for
Br$_{\rm sl}(B)$ does not fall lower than 11.5\%~\cite{Kagan}, while
the corresponding experimental number
is $(10.43\pm0.24)\%~\cite{glasgow}$.
Moreover, no reasonable choice of the parameters above allows one
to eliminate  a very substantial deficit in the charm multiplicity.

Both discrepancies evaporate if the $B$ non-leptonic decays receive a
contribution from the $b\rightarrow s$ + gluon transition,
at the level of $\sim$ 15\% of the total width. Then the theoretical
prediction for  Br$_{\rm sl}(B)$ shifts down to 10.4\%;
simultaneously, the charm multiplicity turns out to be within error
bars.  As was observed in Ref.~\cite{Kagan2}, in supersymmetric
models such a transition can naturally arise, with
the right strength,
if the gluino and squark masses lie in the 100{\rm\,GeV} ballpark.
What is important is that the additional graphs giving rise to
$b\rightarrow s$ + gluon transition do not spoil the $b\rightarrow s$
+ photon transition. Indeed, the ratio of the photon to gluon
probabilities is $(Q_d^2\alpha)/({\alpha_s}\eta^2)$
where $Q_d=1/3$ is the down quark electric charge, and $\eta$
is a numerical factor including, among other effects, an enhancement
of the $b\rightarrow s$ + gluon transition due to the gluon radiative
corrections. According to Ref.~\cite{Kagan2} $\eta\sim 2.5$ to 3.
With ${\alpha_s}\approx 0.11$ the ratio is close to $10^{-3}$. This
means that the $b\rightarrow s$ + gluon transition can well
contribute at the level of 15\%; the corresponding contribution to the
$b\rightarrow s\gamma$ is at the level of $10^{-4}$, which is quite
acceptable phenomenologically~\cite{bsgamma}.

\subsection{$b$--$\tau$ unification}

It has been argued that, in the MSSM alone, with no additional
mass relations, the requirement of strict $b$--$\tau$ mass unification
can only be achieved in a relatively very narrow region of the
(${m_t},\tan\beta$) plane for a wide range of
${\alpha_s(m_Z)}$~\cite{btau,lp2}.
However, it was
noted in Ref.~\cite{kkrw1} that, if ${\alpha_s(m_Z)}$ is small $\sim0.11$,
the above strong relation between $\tan\beta$ and ${m_t}$ can be
significantly relaxed provided that strict unification condition
$h_b/h_{\tau}=1$ at
the GUT scale
is reduced somewhat ($\sim10\%$).
(See Figs.~1 and 2 of Ref.~\cite{kkrw1}.)
GUT-scale uncertainties of this size are actually typically present in
GUT's~\cite{lp2}.

\section{Conclusions}

The observation that the gauge coupling constants, which look so
different at the electroweak scale, evolve and converge at a scale
somewhat smaller than the Planck mass was crucial in the original idea
of grand unification~\cite{gut:original}.
Later on, with more accurate data and more precise calculations
available, it turned out that the gauge couplings do not intersect at
one point. The fact that
we are off by only a relatively very small amount is very encouraging
and shows that the original idea is viable, and only details must be
adjusted. This first led people from the SM to the MSSM. This work
concludes that, if ${\alpha_s(m_Z)}$ is indeed close to $0.11$, the gluino
must be rather light, ${m_{\widetilde g}}\sim100{\rm\,GeV}$, and thus
accessible to
present direct searches. It is also gratifying to note that, with the
mass of the gluino lying in this ballpark, other problems (like the
$R_b$ excess at LEP, a deficit of the semileptonic branching ratio of
B-mesons, {\it etc.}) might find their solutions as well. Finally, many
studies of SUSY, including mass bounds on sparticles and dark matter
searches, rely on the mass
relations~(\ref{monemtwo:eq})--(\ref{mtwomgluino:eq}). This analysis
provides arguments for relaxing them.

\section*{Acknowledgements}

This work was supported in part by the U.S. Department of
Energy under the grant number DE-FG02-94ER40823.




\begin{thebibliography}{99}
\bibitem{F2}
Ref.~\cite{Langacker} quotes even a higher value,
${\alpha_s(m_Z)}=0.127\pm0.005$.
The errors are believed to be dominated by theoretical uncertainties.
The
most exhaustive  theoretical analysis is done for the total hadronic
width
$\Gamma (Z\rightarrow\mbox{hadrons})$. The error quoted above,
$\Delta{\alpha_s(m_Z)}
=\pm 0.005$, is essentially determined by analyzing
$\Gamma (Z\rightarrow\mbox{hadrons})$.
\bibitem{F3}
The following terminology is used throughout the paper:
if ${\alpha_s(m_Z)}$ is 0.12 or larger it will be referred to as
``large ${\alpha_s}$";  if
${\alpha_s(m_Z)}$ is close to 0.11 it will be referred to as
``small ${\alpha_s}$".
\bibitem{F4}
Let us parenthetically
note that some authors investigate a wider range of variations of
${\sin^2\theta_W}(m_Z)$. For instance, in Ref.~\cite{urano} the highest
value considered was ${\sin^2\theta_W}(m_Z)=0.2327$. As was
explained above inching ${\sin^2\theta_W}(m_Z)$ up one lowers
the minimal value of ${\alpha_s(m_Z)}$. Thus, the result of
Ref.~\cite{urano} in the
CMSSM with the Planck mass correction switched off
is ${\alpha_s(m_Z)}_{\rm min}\approx 0.114$. The question whether
the value ${\sin^2\theta_W}(m_Z)=0.2327$ is admissible is left open;
we stick to Eq.~(\ref{s2winput:eq}).
\end{thebibliography}
\end{document}